\documentclass[
 reprint, amsmath,amssymb,
 aps]{revtex4-2}

\usepackage{graphicx}
\usepackage{dcolumn}
\usepackage{bm}
\usepackage{xcolor} 
\usepackage{physics}
\newcommand{\beq} {\begin{equation}}
\newcommand{\eeq} {\end{equation}}
\newcommand{\bea} {\begin{eqnarray}}
\newcommand{\eea} {\end{eqnarray}}

\newcommand{\om}{\omega_m}
\newcommand{\on}{\omega_n}

\newcommand{\bo}{\bar{\omega}}
\DeclareMathOperator{\sgn}{sgn}

\begin{document}

\title{The Effect of Repulsion on Superconductivity at Low Density}

\author{Dan Phan}
\author{Andrey V. Chubukov}
\affiliation{
School of Physics and Astronomy, University of Minnesota, Minneapolis, Minnesota 55455, USA
}

\date{\today}

\begin{abstract}
We examine the effect of repulsion on superconductivity in a three-dimensional system with a Bardeen-Pines-like interaction in the low-density limit, where the chemical potential $\mu$ is much smaller than the phonon frequency $\omega_L$. We parameterize the strength of the repulsion by a dimensionless parameter $f$, and find that the superconducting transition temperature $T_c$ approaches a nonzero value in the $\mu = 0$ limit as long as $f$ is below a certain threshold $f^*$. In this limit, we find that $T_c$ goes to zero as a power of $f^*-f$, in contrast to the high density limit, where $T_c$ goes to zero exponentially quickly as $f$ approaches $f^*$. For all nonzero $f$, the gap function $\Delta (\omega_m)$ changes sign along the Matsubara axis, which allows the system to partially overcome the repulsion at high frequencies. We trace the position of the gap node with $f$ and show that it approaches zero frequency as $f$ approaches $f^*$.
To investigate the robustness of our conclusions, we then go beyond the Bardeen-Pines model and include full dynamical screening of the interaction, finding that $T_c$ still saturates to a non-zero value
at $\mu = 0$ when $f < f^*$.
\end{abstract}

\maketitle

\section{Introduction}
In recent years, there has been a renewed interest in superconductivity at low carrier density, arising from experimental advances in an assortment of materials such as SrTiO$_3$ \cite{Schooley64,Schooley65,Lin14}, single-crystal Bi \cite{Prakash2017,Ruhman2017}, Pb$_{1-x}$Tl$_x$Te~~\cite{Chernik1981}, Bi$_2$Se$_3$ \cite{Kong2013}, and half-Heusler compounds \cite{Nakajima2015} (for recent review see Ref. \cite{gastiasoro2020superconductivity}). As of present, the origin of superconductivity in many of these low-density materials is not well-understood. Due to the dilute nature of these compounds, the repulsive electron-electron interaction is weakly screened, and one naively expects
this repulsion to dominate any attraction arising from the electron-phonon interaction.

Non-$s$-wave superconductivity arising from electron-electron repulsion (e,g., attraction in the $d_{x^2-y^2}$ channel in the cuprates) is well understood as the ultimate result of static screening of the Coulomb repulsion in the particle-hole channel, which leads to Friedel oscillations at large distances and in most cases generates an attraction in one or more non-$s$-wave channels
(see, e.g., Ref. \onlinecite{Maiti2013} and references therein).
However, superconducting order parameters in SrTiO$_3$ and other low-density materials are likely $s$-wave, in which case a static interaction remains repulsive.

A standard formalism, which describes how $s$-wave superconductivity can exist even for a (sufficiently weak)
repulsive interaction, involves dynamical screening
~\cite{Marsiglio_20,Esterlis_20,Berg_20,gastiasoro2020superconductivity}.
 The reasoning is as follows \cite{
 Grabowski1984, Richardson1997,McMillan1968,Tolmachev1958}:
when $\mu$ is much larger than the frequency of the pairing boson, $\omega_L$, the bare Coulomb repulsion is renormalized down by dynamical screening in the particle-particle channel, (the same channel which accounts for superconductivity in the case of attraction), and at frequencies of order $\omega_L$ is reduced by a factor $\ln(\mu/\omega_L)$. If this reduced Coulomb repulsion is smaller than the electron-phonon attraction, the effective interaction is attractive at smaller frequencies, and superconductivity develops. In a more accurate treatment~\cite{Morel1962,Rietschel1983} the full interaction (Coulomb + electron-phonon) remains repulsive, but is reduced at small frequencies. Superconductivity then develops with a frequency-dependent gap, which changes sign between small and large frequencies.  An effective description of a ``conventional'' sign-preserving  $s$-wave superconductor with an attractive interaction (electron-phonon minus Coulomb (reduced by $\ln(\mu/\omega_L)$))
emerges once one integrates out higher-energy fermions, for which the sign of the gap is opposite to that at smaller frequencies.

This reasoning holds in the high-density limit, where $\mu \gg \omega_L$, but is not applicable to the case when the fermionic density is small and $\mu \leq \omega_L$. There are two reasons for this: first, even for an attractive interaction, Cooper pairing is thought to arise from fermions near the Fermi surface, where the density of states can be approximated by its value on the Fermi surface and the pairing kernel is logarithmically singular. As $\mu$ decreases, the range where this description is applicable, shrinks. Keeping contributions from the Cooper logarithm, one then finds that for $\mu \leq \omega_L$, the  prefactor of $T_c$ scales with $\mu$,
leading to the vanishing of $T_c$ as $\mu \to 0$.
Second, by the same smallness of $\mu/\omega_L$, the actual pairing interaction is almost independent of frequency in the range where the density of states is approximated by its value on the Fermi surface, and is repulsive. For such an interaction, there is no solution of the gap equation as there is no way to obtain a sign change in the gap function.

This first argument was re-analyzed by  Gastiasoro et al.~\cite{Gastiasoro2019},  who
solved the full momentum and frequency-dependent Eliashberg equations for $T_c$ in a model of a three-dimensional electron gas with only  the attractive, phonon-mediated  component of the pairing interaction, $V(\Omega,\vb{q}) = V(\vb{q}) \omega^2_L/(\Omega^2 + \omega^2_L)$. They argued that at small $\mu$, typical $q$ for the pairing are much larger than $k_F$. As a result, the prefactor of $T_c$ does not vanish at $\mu =0$. In fact, they found that $T_c$  actually \textit{increases} in the $\mu = 0$ limit and argued that this increase reflects the essentially unscreened nature of the interaction. This result is consistent with the observations made earlier by Takada and others~\cite{Takada1992,Richardson1997,Ruhman2016,Ruhman2017,Tchoumakov2020} that even at moderately-large density, there are features in the gap function away from  $k=k_F$.

In this communication we analyze $T_c$ and the gap structure in the low-density limit within the effective Bardeen-Pines model~\cite{Bardeen1955,Grabowski1984,Gastiasoro2019} with both electron-phonon attraction {\it and} electron-electron repulsion. Specifically, we investigate the pairing of electrons in three dimensions with a spherical Fermi surface, interacting via
 \begin{equation}
    V(\Omega,q) = \frac{4\pi e^2}{q^2 + \kappa^2} \big(f - \frac{\omega_L^2}{\Omega^2 + \omega_L^2} \big).
    \label{eq:nn}
\end{equation}
Here $f$ is a measure of the strength of the repulsive interaction, and $\kappa$ is the Thomas-Fermi wavevector. Gastiasoro et al. considered the case $f=0$, while a physically-motivated interaction corresponds to $f \geq 1$. We  treat $f$ as a parameter and obtain results for values of $f$ both below and above one.

The model of Eq. \ref{eq:nn} excluding the momentum-dependence of the interaction has been studied by a number of authors (see, e.g., Ref. \onlinecite{Chubukov2019} and references therein). The result of these studies is that for a generic $\mu$, superconductivity survives up to some critical $f^* (\mu)$. However, as $\mu$ approaches zero, $T_c$ vanishes for all $f>1$. Our goal is to study how these results are modified if one takes the full momentum and frequency-dependent $V(\Omega,q)$ in  Eq. \ref{eq:nn}.
To this end, we solve the momentum and frequency-dependent integral equations for $T_c$ for $s$-wave pairing. Our calculations show that $T_c$ does approach a finite value at $\mu = 0$, even in the presence of static repulsion. We argue that a non-zero $T_c$ results from including both the sign change of the gap function between small and large frequencies and the fact that at small $\mu$ the pairing involves fermions away from the Fermi surface. As at high density, the pairing holds as long as $f$ is below a certain $f^*(\mu)$. We show that at vanishing $\mu$, $f^*(\mu)$  approaches a finite value $f^*(0) >1$. Unlike the high-density case, where $T_c$ goes to zero exponentially quickly as $f$ approaches the threshold value, we find that in the low-density limit $T_c$ goes to zero as a power law: $T_c \propto (f^* - f)^2$.

The gap function $\Delta(\omega_m)$ at $f < f^*$ changes sign at some $\omega_m = \omega_0$. We show that $\omega_0$
exists for arbitrarily small $f >0$, appearing at infinity when $f = 0^+$. As $f$ increases, $\omega_0$ decreases and ultimately vanishes at $f= f^*$. This result is interesting from a topological perspective, since nodal points of $\Delta (\omega_m)$ correspond to centers of dynamical vortices \cite{Wu2021,Christensen2021}. The nodeless state at $f=0$ and a state with a nodal gap at $f >0$ have different numbers of vortices and are therefore  topologically distinct.  In this respect, the vanishing of  superconductivity at $f = f^*$ can be viewed as a topological transition, when the gap function can no longer hold vortices on the Matsubara axis.

We also obtain $T_c$ as a function of $\mu$ for the more realistic model which includes the full momentum and frequency dependence of the polarization bubble, which accounts for the screening of the Coulomb interaction. We find that $T_c$ remains finite in the limit $\mu \to 0$, as long as $f < f^*$.

The paper is structured as follows: In Section \ref{sec:model}, we introduce and motivate our model.
In Sec.\ref{sec:equations} we present the linearized Eliashberg equations, which we use to calculate $T_c$ and the gap function at $T_c$. In Sec. \ref{sec:rietschel_sham} we briefly review the effect of repulsion on superconductivity at high-density. In Sec. \ref{sec:analytics} we present analytical results in the low-density limit. In Section \ref{sec:numerics} we discuss our numerical analysis.  We review the numerical methods we use to solve the Eliashberg equations in Section \ref{sec:methods}, and  present the results of our calculations in Sec. \ref{sec:results}. In particular, we show (i) how $T_c$ varies with the chemical potential $\mu$, (ii) how $T_c$ is affected by the strength of the repulsive component of the interaction $f$, (iii) how the gap function depends on Matsubara frequency, (iv) how  the location of the nodal point of $\Delta (\omega_m)$ depends on the strength of the repulsive interaction,  and (v) how the gap function depends on momenta away from $k=k_F$. In Sec. \ref{sec:freq_dep} we  analyze $T_c$ and the gap function in a model with dynamically-screened Coulomb interaction.  We present our conclusions in Sec. \ref{sec:conclusions}.

\section{Model}
\label{sec:model}
We consider an electron gas in 3 dimensions, with dispersion $\xi(\mathbf{k}) = k^2/2m - \mu$. Electrons interact via the Coulomb potential and through exchange of lattice vibrations, which effectively screen the electron charge. We follow Ref. \onlinecite{Gastiasoro2019} and approximate the total (direct and phonon-mediated) interaction between electrons by
 \begin{equation}
\label{eq:rpa}
    V(\Omega,q) = \frac{4\pi e^2}{\varepsilon(\Omega)q^2 - 4\pi e^2 \Pi(\Omega,q)},
\end{equation}
 where $\Omega$ is a bosonic Matsubara frequency, $\varepsilon(\Omega)$ is the dielectric function, which incorporates the screening by phonons, and $\Pi(\Omega,q)$ is the electron polarization bubble. We take the dressed dielectric function to be
\begin{equation}
    \varepsilon(\Omega) = \varepsilon_{\infty} \frac{\Omega^2 + \omega_L^2}{\Omega^2 + \omega_T^2},
\label{eq:nn_2}
\end{equation}
where $\omega_L$ and $\omega_T$ are the frequencies of longitudinal and transverse optical phonons, respectively, $\omega_L > \omega_T$, and $\varepsilon_\infty$ is the  dielectric constant in the absence of phonons. In the zero frequency limit, $\varepsilon(0) = \varepsilon_\infty \omega_L^2/\omega_T^2$; this is known as the
Lyddane-Sachs-Teller relation~\cite{Ashcroft1976}. Since the polarization bubble $\Pi (\Omega,q)$ is negative for all $\Omega$ and $q$, the interaction $ V(\Omega,q)$ is positive (repulsive) at all frequencies. The phonons, however, make this interaction frequency-dependent, even if we approximate the polarization bubble by its static, long-wavelength limit $\Pi (\Omega,q) \approx - 2 N(\mu)$. In this approximation,
\begin{equation}
    V(\Omega,q)
    = \frac{4\pi e^2}{\varepsilon(\Omega)q^2 + \kappa^2}
\end{equation}
where $\kappa = (8\pi e^2 N(\mu))^{1/2}$ is the Thomas-Fermi wavevector.

The interaction in the form of Eq. \ref{eq:nn_2} can be rigorously justified for polar insulators, where $\Pi (\Omega,q) =0$ and
\bea
V(\Omega,q) &=& \frac{4\pi e^2}{\varepsilon_{\infty} q^2} ~ \left(1 - \frac{\omega^2_L-\omega^2_T}{\Omega^2 + \omega^2_L}\right) \nonumber \\
&& = \frac{4\pi e^2}{{\tilde \varepsilon}_{\infty} q^2} \left(f -  \frac{\omega^2_L}{\Omega^2 + \omega^2_L}\right)
\label{mo_1}
\eea
where $f = 1/(1-\omega^2_T/\omega^2_L)$ and ${\tilde \varepsilon}_{\infty} = \varepsilon_{\infty} f$.
In Ref. \cite{Ruhman2016} the authors applied a similar model with $f=1$
  to a polar crystal with a finite density of conduction electrons, appropriate for SrTiO$_3$.  Eq. \ref{eq:nn_2} can also be justified for a
non-polar crystal with a monoatomic basis, where $\varepsilon_{\infty} =1$ and $\omega_T =0$ as there are no transverse phonons. In this case, at finite electron density, we have
\begin{equation}
    V^\text{BP}(\Omega,q) = \frac{4\pi e^2}{q^2 + \kappa^2}
    \bigg(1 - \frac{\omega^2_q}{\Omega^2 + \omega^2_q} \bigg),
\label{mo_2}
\end{equation}
where $\omega_q  = q\omega_L/\sqrt{q^2 + \kappa^2}$ is the phonon frequency. This is known as  the Bardeen-Pines model. The frequency $\omega_q \approx \omega_L$ at $q \gg \kappa$ and becomes linear in $q$ for $q \ll \kappa$ due to electronic screening.

For most of the paper, we follow Ref.~\cite{Gastiasoro2019} and use the semi-phenomenological form of $V(\Omega,q)$:
\begin{equation}
    V(\Omega,q) = \frac{4\pi e^2}{q^2 + \kappa^2} \big(f - \frac{\omega_L^2}{\Omega^2 + \omega_L^2} \big).
    \label{eq:interaction}
\end{equation}
We set $\omega_L$ to be a constant ($\omega_L = 0.1$eV) and treat $f$  as a parameter, which we vary. $ V(\Omega,q)$  interpolates between Eqs. (\ref{mo_1}) and (\ref{mo_2}) and can be thought of as an extended Bardeen-Pines model. We keep $\kappa$ finite, but will chiefly focus on the low-density limit, where typical $q$ are much larger then $\kappa$. In this situation, $f=1$ corresponds to the Bardeen-Pines interaction for a non-polar crystal, while for $f > 1$ the interaction closely mirrors the electron-electron interaction in a polar crystal.

Keeping $f$ as a parameter will also allow us to connect to previous work~\cite{Gastiasoro2019}, which considered
Eq. (\ref{eq:nn}) in the purely-attractive $f = 0$ limit.  We later extend the model by replacing $\kappa^2$ with $-4 \pi e^2 \Pi ( \Omega,q)$ and show that the key results, obtained with Eq. \ref{eq:interaction}, survive.

\subsection{Equations for the fermionic self-energy and the pairing vertex}
\label{sec:equations}

The interaction $ V(\Omega,q)$ gives rise to corrections to the fermionic dispersion and the fermionic residue, while also mediating pairing between fermions.  We assume that the fermionic self-energy  can be evaluated in the one-loop approximation and the pairing vertex can be evaluated in the ladder approximation, both using dressed Green's functions for the intermediate fermions. These approximations amount to neglecting vertex corrections to the interaction. At large density ($\mu/\omega_L \gg 1$), such approximations can be justified by invoking Migdal's theorem \cite{Migdal1958}.
However, for $\mu \leq \omega_L$, there is no rigorous justification for neglecting vertex corrections. The authors of Ref. \onlinecite{Gastiasoro2019} argued that for $f=0$, vertex corrections are of order one and do not affect the results qualitatively. We assume that this holds also for finite $f$.

Neglecting vertex corrections, we obtain a set of three coupled equations for the inverse quasiparticle residue $Z_n (\varepsilon)$, the pairing vertex $\phi_n (\varepsilon)$, and the correction to fermionic dispersion $\chi_n(\varepsilon)$.
Here the index $n$ refers to Matsubara frequency $\omega_n = (2n+1)\pi T$ and $\varepsilon$ is the quasiparticle dispersion $\varepsilon_k = k^2/2m$. The two variables $n$ and $\varepsilon$ parameterize the frequency and momentum dependence of the residue and the correction to the dispersion and of the pairing vertex. We consider only $s$-wave pairing, where the pairing vertex has no angular dependence, and focus on $T =T_c$, where the pairing vertex is infinitesimally small. The three equations are

\begin{widetext}
\begin{equation}
Z_n(\varepsilon) - 1 = - T \frac{1}{\on}\sum_m \int_0^\infty d\varepsilon' N(\varepsilon')V_{n-m}^\mathrm{s-wave}(\varepsilon,\varepsilon') \frac{\om Z_m(\varepsilon')}{[\om Z_m(\varepsilon')]^2 + [\varepsilon'-\mu + \chi_m(\varepsilon')]^2}
 \label{eq:Z}
\end{equation}
\begin{equation}
\phi_n(\varepsilon) = - T \sum_m \int_0^\infty d\varepsilon' N(\varepsilon') V_{n-m}^\mathrm{s-wave}(\varepsilon,\varepsilon') \frac{\phi_m(\varepsilon')}{[\om Z_m(\varepsilon')]^2 + [\varepsilon'-\mu + \chi_m(\varepsilon')]^2}
 \label{eq:phi}
\end{equation}
\begin{equation}
\chi_n(\varepsilon) = T \sum_m \int_0^\infty d\varepsilon' N(\varepsilon')V_{n-m}^\mathrm{s-wave}(\varepsilon,\varepsilon') \frac{\chi_m(\varepsilon') + \varepsilon'-\mu}{[\om Z_m(\varepsilon')]^2 + [\varepsilon'-\mu + \chi_m(\varepsilon')]^2},
 \label{eq:chi}
\end{equation}
\end{widetext}

and the gap function is given by $\Delta_n(\varepsilon) = \phi_n(\varepsilon)/Z_n(\varepsilon)$.

In our modified Bardeen-Pines model, we obtain
\begin{align}
    V_{n-m}^\mathrm{s-wave}(\varepsilon_k,\varepsilon_q)
    &= \int_{-1}^1 \frac{d\cos\theta}{2}V_{n-m}(\sqrt{k^2 + q^2-2kq\cos\theta})\\
    &= \frac{\pi e^2}{kq}\ln\bigg( \frac{(k+q)^2+\kappa^2}{(k-q)^2+\kappa^2}\bigg) u_{n-m},
\label{eq:nn_9}
\end{align}
where we have defined $u_{n} = f - \omega_L^2/(\on^2+\omega_L^2)$. For notational convenience, we will drop the \textit{s}-wave
 super-script
  on the interaction.

\subsubsection{Migdal-Eliashberg Approximation}
\label{sec:approx}

In the Migdal-Eliashberg approximation, the  integrals over the dispersion (i.e., over $\varepsilon'$)  are evaluated by linearizing the integrand about the chemical potential $\mu$. In this approximation, the variation of the density of states $N(\varepsilon)$ and the interaction $V_{n-m}(\varepsilon,\varepsilon')$ with $\varepsilon$, $\varepsilon'$  is ignored, and both quantities are evaluated at $\varepsilon = \mu$. With this approximation, the integrals over energy can be evaluated. One then finds that $\chi_n (\varepsilon) =0$, while $Z_n \equiv Z_n (\mu)$ and $\phi_n  \equiv \phi_n(\mu)$ only depend on Matsubara frequency:

\begin{equation}
    Z_n = 1-\pi N(\mu)  \frac{T}{\on}\sum_m V_{n-m} \sgn(\om)
\end{equation}
\begin{equation}
    \phi_n = - \pi N(\mu) T \sum_m V_{n-m} \frac{\phi_m}{\abs{\om}Z_m}.
\end{equation}

These two equations are known as the Eliashberg equations. The approximation $\varepsilon, \varepsilon' \approx \mu$ used above to obtain these equations, is valid if pairing only involves fermions near the Fermi surface. This holds in the adiabatic limit where $\mu \gg \omega_L$, but not when $\mu \leq \omega_L$. Since we are concerned with the low-density limit, we instead work with the full set of equations, Eqs. (\ref{eq:Z}-\ref{eq:chi}).

\subsection{The High-Density Limit}
\label{sec:rietschel_sham}

Before we proceed, we briefly review what is known in the high-density limit, $\mu \gg \omega_L$.  In this limit, typical $q$ are large and of order $k_F$. To first approximation, the momentum dependence of the interaction can then be approximated by
$V_0 \sim e^2/k^2_F$, and the full interaction can be
approximated as $V_0 (f - \omega^2_L/(\Omega^2 + \omega^2_L))$.  At weak coupling, when $\lambda = N (\mu) V_0 \sim r_s$ is small, the fermionic self-energy can be neglected, and the Eliashberg equation for the pairing vertex can be solved with $Z_n(\varepsilon) = 1$. In this case, $T_c$ is finite for $f$ below a certain cutoff $f_c >1$ which depends on $\lambda$ \cite{Chubukov2019}. In other words, Cooper pairing continues to exist even if the interaction is purely repulsive ($f > 1$), as long as the repulsion is sufficiently weak ($f < f_c$).  However, for any $f >0$, the gap function changes sign as a function of Matsubara frequency. For $f >1$, this sign change allows the system to partially
neutralize the ``average'' repulsion and gain from reduction of the repulsion at small frequencies \footnote{This is analogous to the purely momentum-dependent case, where the gap changes sign as a function of momentum to orthogonalize against a repulsive s-wave component of the interaction and take advantage of an attractive component in a different angular-momentum channel \cite{Maiti2013}.}.

An illustrative toy model highlighting these points was introduced by Rietschel and Sham \cite{Rietschel1983}. It mirrors the frequency dependence of the modified Bardeen-Pines model:

\begin{equation}
    V_{mn}
    =
    \begin{cases}
    0, \quad \abs{\om} > E_c \quad \text{or} \quad \abs{\on} > E_c\\
    U(f-\Theta(\omega_L-\abs{\om})\Theta(\omega_L-\abs{\on})), & \text{otherwise.}
    \end{cases}
\end{equation}

In this model, electrons experience a repulsion of magnitude $Uf$ for all frequencies below some cutoff $E_c$. However, if both electrons have frequency smaller than $\omega_L$, there is an additional attractive term $-U$. Using this interaction and assuming weak coupling, $U N (\mu) \ll 1$, one can show that $T_c$ comes from fermions in the vicinity of the Fermi level and is given by

\begin{equation}
    T_c = 1.13 \omega_L e^{-1/\lambda},
\end{equation}
where
\begin{equation}
\label{eq:coupling_RS}
    \lambda = N(\mu) U \bigg(1 - \frac{f}{1 + f N(\mu) U \ln(1.13 E_c/\omega_L)} \bigg).
\end{equation}

We see that an increase in either $N(\mu)$ or $U$ enhances $T_c$. In particular, by increasing $N(\mu)$ or $U$, we increase the prefactor in Eq. \ref{eq:coupling_RS} and further reduce the repulsive contribution. Conversely, if either $N(\mu)$ or $U$ is reduced, the effective repulsive contribution is enhanced relative to the attractive term. Superconductivity exists as long as $f < f_c$, where

\begin{equation}
    f_c = \frac{1}{1 - N(\mu) U \ln(1.13 E_c/\omega_L)}  >1
\end{equation}
Note that for $f \leq 1$, superconductivity exists for arbitrarily small values  of $U N(\mu)$ and $T_c \propto \exp(-1/N(\mu) U)$. If $f = 1$, $\lambda \propto (N(\mu) U)^2$, so that $T_c \propto \exp(-1/(N(\mu) U)^2)$. This behavior is also seen in the model with $V(\Omega) \propto \left(f - \frac{\omega_L^2}{\Omega^2 + \omega_L^2} \right)$~\cite{Chubukov2019}.

The gap function in the Rietschel-Sham model has a low-frequency component $\Delta_1$ and a high-frequency component $\Delta_2$. The two are of opposite sign, and are related by

\begin{equation}
\Delta_2 = -\frac{N(\mu) U f}{1+f N(\mu) U \ln(1.13 E_c/\omega_L)} \ln(\frac{1.13 \omega_L}{T_c})\Delta_1.
\end{equation}

As $f$ increases towards $f_c$, $T_c$ decreases. From the above formula, this implies that the high-frequency gap becomes more and more negative, with the ratio $\Delta_2/\Delta_1$ diverging as $T_c$ goes to zero. However, the position of the sign change of the $\Delta(\om)$ is fixed at $\omega_0 = \omega_L$. This is a consequence of the fact that the boundary between low-frequency and high-frequency regimes in the Rietschel-Sham model is fixed at $\omega_L$.  In the model with $V(\Omega) \propto \left(f - \frac{\omega_L^2}{\Omega^2 + \omega_L^2} \right)$, the position of the zero of $\Delta (\omega_m)$ at $\omega_m = \omega_0$  is set by the solution of the gap equation and varies with $f$.
In the high-density limit ($\mu \gg \omega_L$), $\omega_0$ is nearly infinite at infinitesimally small $f$ and tends to zero as $f$ approaches the critical $f_c(\mu)$ from below \cite{Pimenov}. We show below that the same holds in the low-density limit.

\section{Analytical results in the Low Density Limit}
\label{sec:analytics}

In our analytical study we follow Refs. ~\cite{Gastiasoro2019,Chubukov2019,Pimenov}. Taking $\mu \to 0$, our goal is to find the critical $f^*$ where $T_c$ vanishes, the relation between $T_c$ and $f^*-f$ (in the limit where $T_c \ll \omega_L$ and $f^* -f \ll 1$), the relation between the position of the gap node $\omega_0$ and $f^*-f$, and the frequency dependence of the gap near $f^*$. Note that we use $f^*$ instead of $f_c$ to distinguish between the low and high-density limits.

For nonzero $\omega_L$, the system is in the Fermi liquid regime, implying that the inverse quasiparticle residue $Z_n (\varepsilon)$ tends to a constant at small frequencies and approaches $1$ at large frequencies, while the correction to the dispersion $\chi_n (\varepsilon)$ is non-singular.
 For an order-of-magnitude analysis, we
 can then set $Z_n (\varepsilon) =1$ and neglect $\chi_n (\varepsilon)$.  Eq. \ref{eq:phi} for the pairing vertex $\phi_n (\varepsilon)$ is then essentially the gap equation.  Introducing
 $p =\sqrt{2m \varepsilon}$ and re-scaling all variables by $\omega_L$ as
 ${\bar T} = T/\omega_L$, $\bar{p} = p/\sqrt{2 m \omega_L}$,
 we re-express Eq. \ref{eq:phi}  as
\begin{eqnarray}
\label{eq:gapeq_neqn_n0}
  {\phi}_n({\bar p})&=& -{\bar T} \sum_m \frac{2\sqrt{\bar \rho}}{\pi}
  \left(f-\frac{1}{1+({\bar \omega}_n-{\bar \omega}_m)^2}\right) \nonumber \\
  && \times \int_0^{\infty} d\bar p'  \frac{\frac{\bar{p}'}{\bar{p}} \log \left(\frac{\bar p+ \bar p'}{|\bar p-\bar p'|}\right)}{\bar p^{'4} + {\bar \omega}^2_m} { \phi}_m (\bar p'),
\end{eqnarray}
where we have introduced $\bar \rho = \mathrm{Ry}/\omega_L$ and $\mathrm{Ry} = m e^4/2= 13.6$ eV is the Rydberg energy. One can show that $\phi_m(\bar{p})$ is independent of $\bar{p}$ for $\bar{p}\ll 1$ and decays as $1/\bar{p}^2$ for $\bar{p}\gg 1$. Setting $\bar{p}\ll 1$ and expanding the logarithm in ${\bar p}$, we find
\begin{eqnarray}
\label{eq:gapeq_neqn_n1}
  {\phi}_n&=& -{\bar T} \sum_m \frac{4\sqrt{\bar \rho}}{\pi}
  \left(f-\frac{1}{1+({\bar \omega}_n-{\bar \omega}_m)^2}\right) \nonumber \\
  && \times \int_0^{\infty} d\bar p' \frac{1}{\bar p^{'4} + {\bar \omega}^2_m} { \phi}_m (\bar p'),
\end{eqnarray}
where we use as shorthand $\phi_n \equiv \phi_n(\bar{p}=0)$.
Since the majority of the weight in the $\bar{p}'$ integral comes from $\bar{p}' \sim |{\bar \omega}_m|^{1/2} \leq 1$,
we can replace $\phi_m(\bar{p}')$ with $\phi_m(\bar{p}'=0)$ on the right-hand-side. Integrating then over $\bar{p}'$, we obtain
 \begin{align}
\label{eq:gapeq_neqn_n2}
  \phi_n & = - g \pi  {\bar T}  \sum_m
  \left(f-\frac{1}{1+({\bar \omega}_n-{\bar \omega}_m)^2}\right) \frac{\phi_m}{|{\bar \omega}_m|^{3/2}},
\end{align}
 where $ g = (2\bar \rho)^{1/2}/\pi$.

To analyze the structure of $ \phi_n =  \phi (\omega_n)$ it is convenient to replace the sum over Matsubara frequencies by an integral and set  the lower cutoff of the integral over $\bo'$ at $O({\bar T})$. Doing so, we obtain
\begin{eqnarray}
 && \phi (\bo_n)   = -g  \int_{O(\bar{T})}^\infty d \bo_m \times \label{eq:gapeq_neqn_n22} \\
 &&
  \left(f-\frac{1}{2} \left(\frac{1}{1+(\bo_n -\bo_m)^2} + \frac{1}{1+(\bo_n +\bo_m)^2}\right)\right)
 \frac{\phi (\bo_m)}{(\bo_m)^{3/2}} \nonumber.
\end{eqnarray}
From this equation, we have
 \begin{align}
\label{eq:gapeq_neqn_n3}
  \phi (0)  & = -g \int_{O(\bar{T})}^\infty d \bo_m
  \left(f-\frac{1}{1+\bo_m^2}\right)
 \frac{\phi (\bo_m)}{(\bo_m)^{3/2}}
\end{align}
 and
  \begin{align}
\label{eq:gapeq_neqn_n4}
  \phi (\bo_n)  & = \phi (0) \left(1 - g Q \frac{\bo_n^2}{1 + \bo_n^2}\right)  + ...,
 \end{align}
where $Q = \int_{O(\bar{T})}^\infty d \bo'/(\bo')^{3/2} \sim 1/{\bar T}^{1/2}$ and the unwritten terms account for $O(g)$ corrections, which are irrelevant for $g \leq 1$. Substituting $\phi (\bo_n)$ from Eq. \ref{eq:gapeq_neqn_n4} into Eq. \ref{eq:gapeq_neqn_n3}, we obtain the following self-consistent equation for $\bar{T}_c$:
\beq
\frac{1-\beta g}{1-\alpha g} -f =  f^* - f = \frac{1}{g Q (1-\alpha g)}.
\label{eq:gapeq_neqn_n5}
\eeq
Here, $\alpha = \int_0^\infty d\bo' (\bo')^{1/2}/(1 + (\bo')^2) = \pi/\sqrt{2} \approx 2.22$ and $\beta = \int_0^\infty d\bo' (\bo')^{1/2}/(1 + (\bo')^2)^2 = \pi/(4\sqrt{2}) \approx 0.56$. Note that $f^* > 1$, since $\alpha > \beta$.
Using
  $Q \sim {\bar T}^{1/2}$, we find the scaling relation
  \beq
  {\bar T}_c \sim  (f^*-f)^2.
   \label{eq:gapeq_neqn_n6}
 \eeq
Next, from Eq. \ref{eq:gapeq_neqn_n4} we see that at large $Q$, i.e small ${\bar T}_c$, the frequency $\bo_0$ at which
$ \phi (\bo_n)$ changes sign, is
\beq
\bo_0 \approx \frac{1}{(g Q)^{1/2}} = (f^*-f)^{1/2}.
  \label{eq:gapeq_neqn_n7}
 \eeq
For  smaller $f$, this expression extends to $\bo_0 \sim ((f^*-f)/f)^{1/2}$. Since $\bar{\omega}_0$ tends to zero as $f$ approaches $f^*$, the gap $\phi(\bar{\omega}_n)$ changes sign at progressively smaller $\bar{\omega}_0$. The vanishing of $\omega_0$ as $f$ approaches $f^*$ is consistent with the behavior near $f_c$ in the high-density limit~\cite{Pimenov}. However, the power-law behavior of $T_c$ and the relation $\omega_0 \sim \left(T_c\right)^{1/4}$ are specific to the case of low-density.

\begin{figure}
\includegraphics[width=\columnwidth]{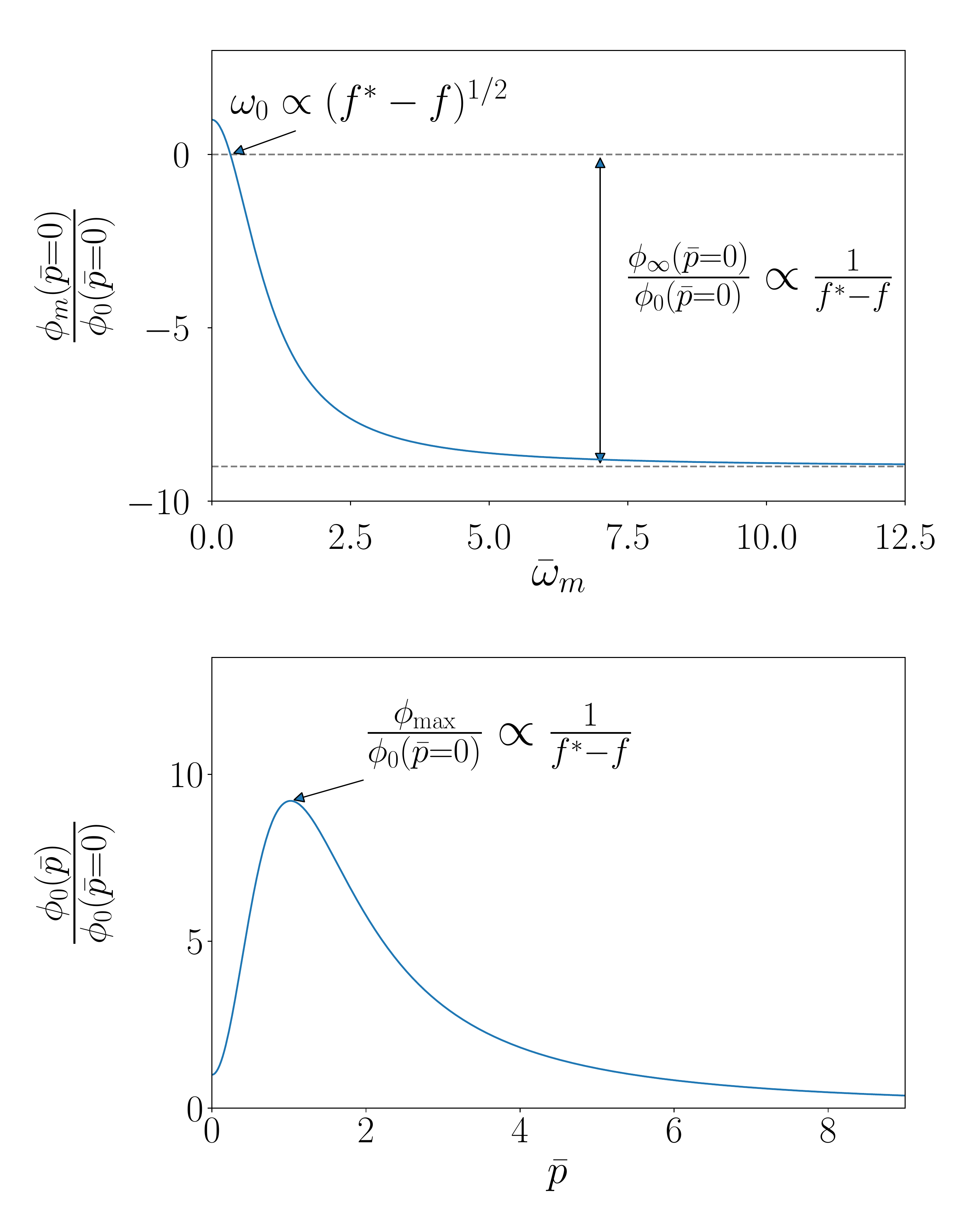}
\caption{\label{fig:gap_schematic} The qualitative behavior of the gap function $\phi_m(\bar{p})$ at $\mu = 0$ as a function of $\bo_m$ in the top panel and as a function of $\bar{p}$ in bottom panel). In addition, we highlight how the gap node $\omega_0$, the high-frequency gap ratio $\phi_\infty(\bar{p}=\infty)/\phi_0(\bar{p} = 0)$, and the peak in $\phi_0(\bar{p})/\phi_0(\bar{p}=0)$ scale near $f^*-f$.}
\end{figure}

Because  $\phi(\bar{\omega}_n) = \phi(0)(1-gQ\bar{\omega}_n^2/(1+\bar{\omega}_n^2))$ and $Q \propto 1/(f^*-f)$, the ratio $\phi(\bar{\omega}_n)/\phi(0)$ becomes more negative with increasing $\bar{\omega}_n$, going as $\phi(\bar{\omega}_n \gg 1)/\phi(0) \propto 1/\bar{T}_c^{1/2}$. This  is shown in the top panel of Fig. \ref{fig:gap_schematic}.

Thus far, we have considered only the $\bar{p}\ll 1$ limit for $\phi_n(\bar{p})$, since this is sufficient to obtain $\bar{T}_c$. We now examine the behavior of $\phi_n(\bar{p})$ at $\bar{p} \geq 1$.
 At large ${\bar p}$, one can easily verify that $\phi_n(\bar{p}) \propto 1/{\bar p}^2$.
Accordingly, we introduce the function $B_n$ via $\phi_n(\bar{p} \gg 1) = B_n/\bar{p}^2$.
Substituting this into Eq. \ref{eq:gapeq_neqn_n0}, taking $\bar{p}\gg 1$, and setting $n=0$ for definiteness, we obtain

\begin{align}
  B_0
  &\approx
  -\bar{T} \sum_m \frac{4\sqrt{\bar{\rho}}}{\pi}
  \left(f-\frac{1}{1+{\bar \omega}^2_m}\right) \nonumber \\
  & \times
  \bigg(\phi_m\int_0^1 d\bar p' \frac{\bar{p}^{'2}}{\bar{p}^{'4} + \bar{\omega}^2_m}
  +
  B_m\int_1^{\bar p} d\bar p' \frac{\bar{p}^{'2}}{\bar{p}^{'4} + \bar{\omega}^2_m} \nonumber \\
  &+ B_m \int_{\bar p}^{\infty} d\bar p' \left(\frac{{\bar p}}{{\bar p}'}\right)^2 \frac{1}{\bar{p}^{'4} + \bar{\omega}^2_m}\bigg).
\end{align}

To obtain the first term of the second line,
 we use our earlier result that for $\bar{p}\ll 1$, $\phi_m(\bar{p})$ is independent of momentum and
 replace $\phi_m(\bar{p})$ with $\phi_m \equiv \phi_m(\bar{p}=0)$ for $\bar{p} < 1$. Similarly, we replace $\phi_m(\bar{p})$ with its asymptotic limit $B_n/\bar{p}^2$ for $\bar{p}> 1$ to obtain the latter two terms of the above equation. To simplify this equation, we note that the last term on the right-hand side is smaller than the first two terms by a factor of order $O(1/\bar{p}^3)$ and is therefore irrelevant. Discarding this term and taking $\bar{p}\to \infty$ in the upper limit of the second integral, we then find

\begin{align}
  B_0
  &\approx
  -\bar{T} \sum_m \frac{4\sqrt{\bar{\rho}}}{\pi}
  \left(f-\frac{1}{1+{\bar \omega}^2_m}\right) \nonumber \\
  & \times
  \bigg(\phi_m\int_0^1 d\bar p' \frac{\bar{p}^{'2}}{\bar{p}^{'4} + \bar{\omega}^2_m}
  +
  B_m\int_1^{\infty} d\bar p' \frac{\bar{p}^{'2}}{\bar{p}^{'4} + \bar{\omega}^2_m}
  \bigg)
  \label{eq:largep}
\end{align}

One can verify that all momentum and frequency integrals appearing in this equation are $O(1)$, i.e. nonsingular in the $T=0$ limit. Assuming that the primary contribution to $B_0$ on the right-hand side of the above equation comes from $\phi_m$, not $B_m$, we find that $B_0$ is determined by $\phi_m$ at frequencies ${\bar \omega}_m = O(1)$, where  $\phi_m \sim -gQ\phi_0$.

We then have $B_0 \sim g^2Q\phi_0$, which, recalling that $Q \sim 1/T_c^{1/2}$,
implies that $B_0/\phi_0$ diverges as $T_c \to 0$ \footnote{One can verify that our assumption (that $\phi_m$ matters much more than $B_m$) is valid by inserting our result, $B_m \sim g\phi_m$ back into Eq. \ref{eq:largep}.}.
 In other words, the ratio $\phi_0(\bar{p})/\phi_0 (\bar{p}=0)$ must have a large peak as a function of $\bar{p}$, with its magnitude growing as $1/(f^*-f)$. We show this behavior in $\phi_0(\bar{p})/\phi_0 (\bar{p}=0)$ in the bottom panel of Fig. \ref{fig:gap_schematic}.

Note also that $T_c$ is finite even if one does not impose an ultraviolet cutoff on the frequency integration.  This is due to the $1/q^2$ momentum dependence of the interaction, which leads to the momentum integration in the particle-particle bubble going as $1/\omega^{3/2}$. In this case, the integral  $\int d \omega/\omega^{3/2}$ converges in the ultraviolet.

\section{Numerical analysis}
\label{sec:numerics}

\subsection{Methods}
\label{sec:methods}
To solve the linearized Eliashberg equations for $T_c$, we note that the equation for the pairing vertex $\phi_n(\varepsilon)$ is essentially an eigenvalue problem. To solve this eigenvalue problem,  we create a linear operator mapping $\phi_m(\varepsilon')$ to $\phi_n(\varepsilon)$ in Eq. \ref{eq:phi} for different values of temperature. To find $T_c$, one must then find at what
 temperature this linear operator's largest positive eigenvalue is equal to 1.

To construct this operator at a given temperature $T$, we first solve for $Z_n(\varepsilon)$ and $\chi_n(\varepsilon)$. This is done self-consistently by iterating Eq. \ref{eq:Z}) and Eq. \ref{eq:chi} starting from $Z_n^\text{initial}(\varepsilon) = 1$ and $\chi_n^\text{initial}(\varepsilon) = 0$ until convergence is reached. The energy integrals are obtained using upper cutoffs from $\Lambda = 100\omega_L$ to $\Lambda=200\omega_L$, and a grid of hundreds of sampling points. We split the energy range into 3 regions, (i) $\varepsilon < \mu - \delta$, (ii) $\mu - \delta < \varepsilon < \mu + \delta$, and (iii) $\mu + \delta < \varepsilon < \Lambda$, where we take $\delta = \mu / 100$. In region (ii) near the chemical potential, we use a high density of quadrature points to account for the peak in the integrand and apply the trapezoidal rule. In regions (i) and (iii) where the variation in the integrand is smoother, we use Gauss-Legendre quadrature to calculate the integrals. In the low-density limit $\mu = 0$, we use a composite Gauss-Legendre grid with around 1000 points, to ensure that we accurately obtain contributions from all values of $\varepsilon$. We find that trends in $T_c$ are well-converged with respect to variations in $\delta$ and the number of quadrature points.

To calculate the Matsubara sums, we note that all Matsubara sums appearing in the Eliashberg equations are convolutions.
These convolutions can be efficiently calculated by first transforming to imaginary time, where the convolution becomes point-wise multiplication. After point-wise multiplication, one can then transform back to Matsubara frequency. When using the Fast Fourier transform, this method scales significantly better, $O(N\log(N))$, than naively calculating the sums directly in Matsubara space, $O(N^2)$.

Though this method works well for larger values of $T_c$, we run into memory issues when trying to extend this method to
temperatures smaller than $T_c \sim 10^{-4}\omega_L$. We obtained  $T_c$  in this temperature range by extrapolating from higher temperature data using the implicit renormalization method \cite{Chubukov2019}, which also allows us to infer the gap function at $T_c$ from less memory-intensive high-temperature calculations. To apply this method, we divide Eqs. (\ref{eq:Z}-\ref{eq:chi}) into low-energy and high-energy components. We then use the high-energy components to obtain an effective gap equation for the low-energy component of the gap. A gap component is considered high-energy if its respective momentum and frequency satisfy $\sqrt{(k^2/2m-\mu)^2 + \omega_m^2} > \Omega_c$, where $\Omega_c$ is some cutoff frequency. For consistency, we take $\Omega_c = \omega_L$ in all calculations. If the largest eigenvalue obtained from this effective gap equation scales linearly with $\ln(\omega_L/T)$, then we can extrapolate $T_c$ by extrapolating the eigenvalues of less-computationally-intensive, high-temperature calculations.

We find that this method captures the overall trend of $T_c$ relatively well. In particular, the transition temperatures calculated from both the traditional eigenvalue method and the implicit renormalization method agree well for large temperatures, where the traditional eigenvalue method is practicable, and for sufficiently large $\mu$.
In the small-$\mu$ limit, we find that the trend in $T_c$ found via the implicit renormalization method is very similar to that found using the eigenvalue method. However, due to the shortcomings of this method in the low-density limit, we calculate $T_c$ and other quantities using the standard eigenvalue method at low density when possible.

Lastly, we find in our calculations that the function $\chi_n(\varepsilon)$ does not vary significantly with momentum or frequency, and can be largely absorbed into the definition of the chemical potential.
Therefore, we expect all results to be relatively insensitive to whether $\chi_n (\varepsilon)$ is included or set to zero. As such, we take $\chi_n(\varepsilon) = 0$ in the following calculations and solve only the two coupled equations for $Z_n(\varepsilon)$ and $\phi_n(\varepsilon)$.

\subsection{Results}
\label{sec:results}

\subsubsection{$T_c$ vs. $\mu$ and $f$}

\begin{figure}
\includegraphics[width=\columnwidth]{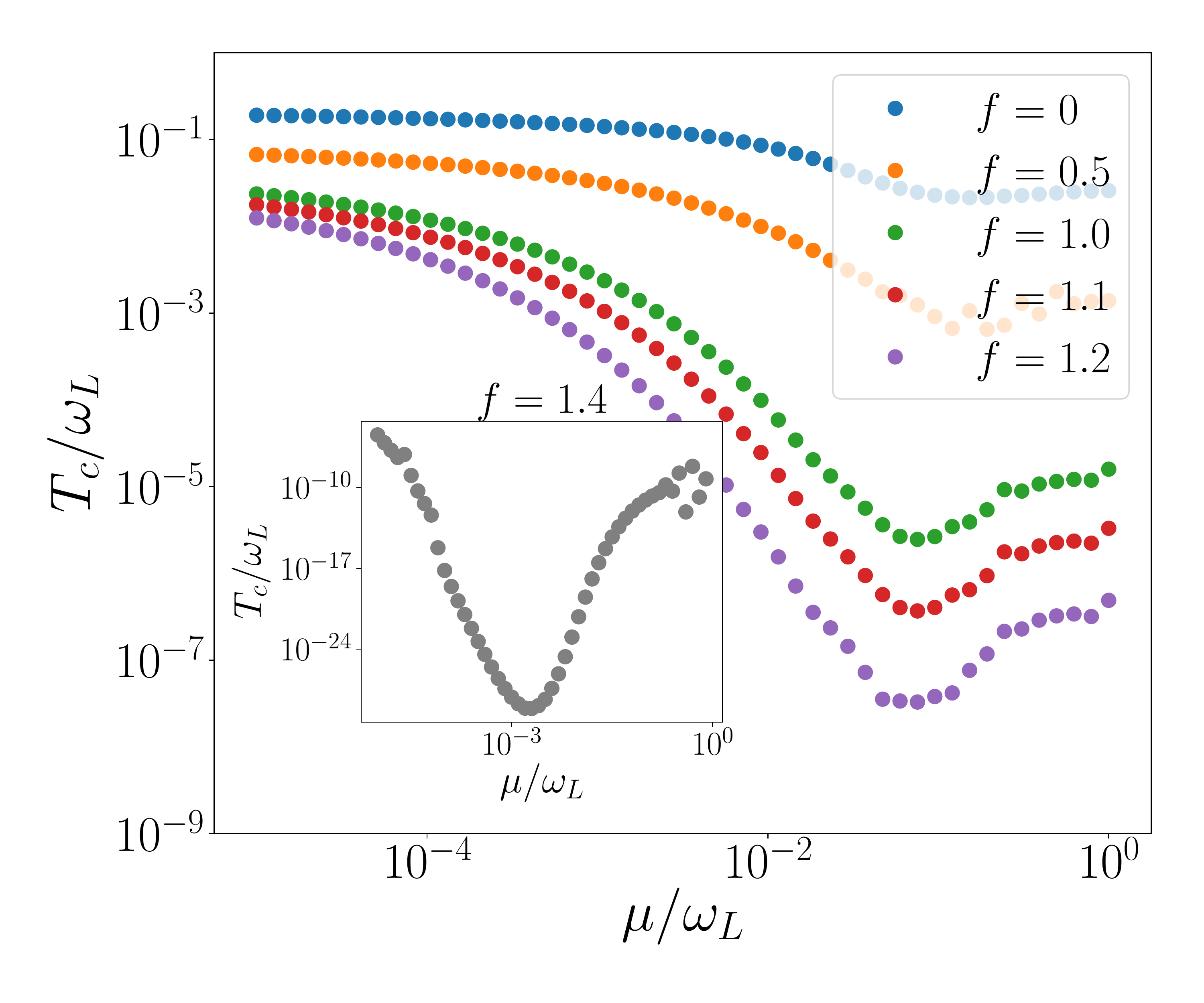}
\caption{\label{fig:tcvsmu}$T_c$ as a function of chemical potential $\mu$ for different strengths $f$ of the repulsive interaction.}
\end{figure}

In Fig. \ref{fig:tcvsmu}, we show how $T_c$ varies with the chemical potential $\mu$ for different values of the repulsive term $f$. For consistency, we use the implicit-renormalization method to extract $T_c$ for all values of $f$ and $\mu$ presented here. We see that $T_c$ is enhanced as one approaches $\mu = 0$, regardless of the strength of the repulsive term. One can understand this enhancement in the same way as was done in the purely attractive case~
 \cite{Gastiasoro2019}. Namely, as $\mu$ decreases, $N(\mu) V_{n-m}(\mu,\mu)$ is enhanced at low density due to the reduction in screening. Additionally, pairing is no longer restricted to occur in a narrow window around $\mu$.
We note that a similar trend in $T_c$ as a function of density has been found by Takada \cite{Takada1992}, who solved the full Eliashberg equations in a multi-valley electron gas to study plasmon-induced superconductivity.

\begin{figure}
\includegraphics[width=\columnwidth]{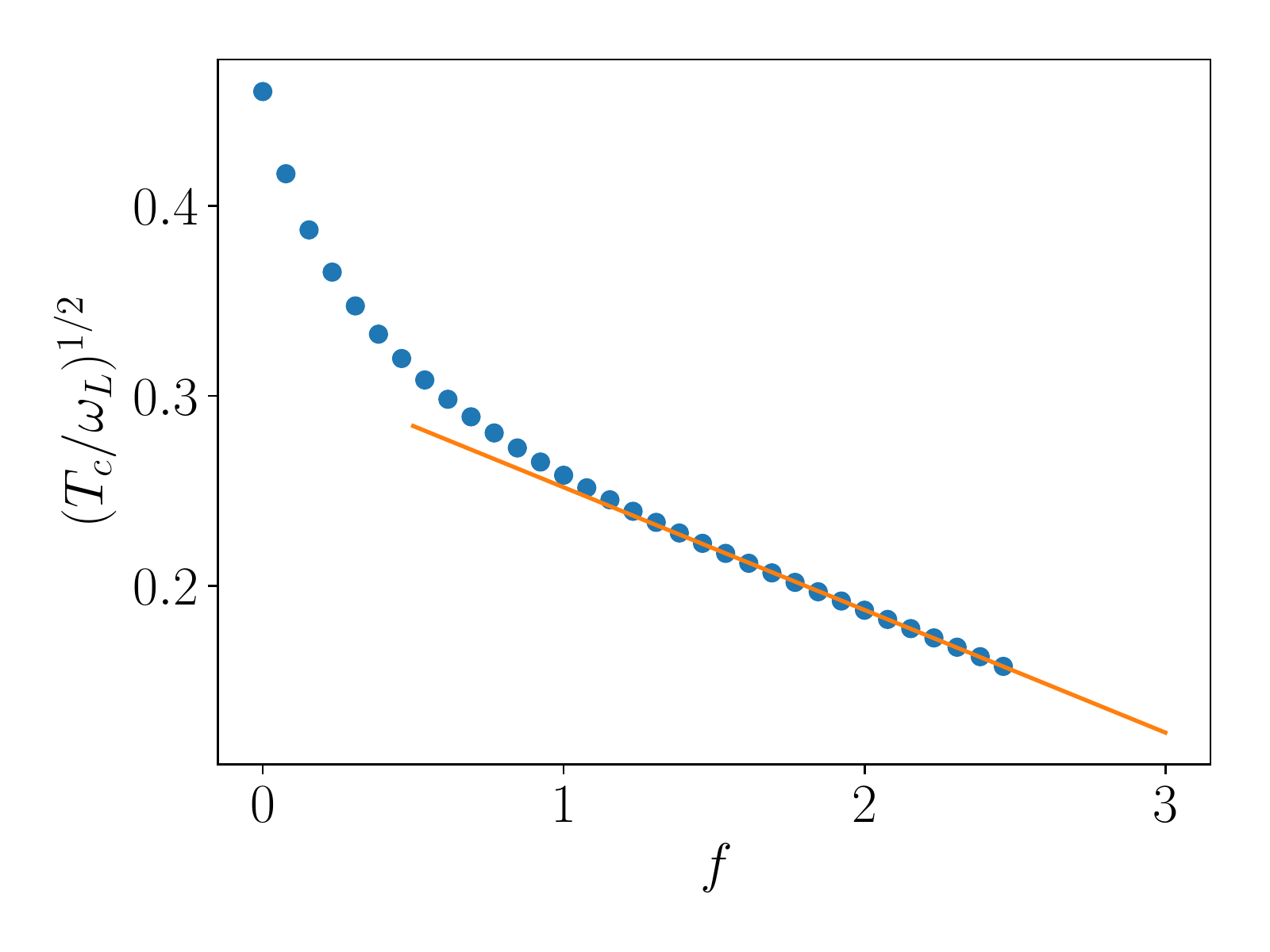}
\caption{\label{fig:scaling} The $\mu = 0$ scaling of $T_c$ as $f$ approaches $f^*$. Overlaid is the line showing that $T_c \propto (f^*-f)^2$ near $f=f^*$.}
\end{figure}

Another trend we find in Fig. \ref{fig:tcvsmu} is that $T_c$ drops more precipitously with increasing $\mu$, passing through a local minimum as a function of $\mu/\omega_L$. The presence of this minimum can understood as follows: as $\mu$ increases from $0$, the effective interaction decreases and the relevant values of $\varepsilon$ gradually cluster closer to $\varepsilon = \mu$. Both factors lead to a reduction of $T_c$ as $\mu$ is increased. However, as $\mu$ increases, the range in $\varepsilon$ about which $1/(Z^2\om^2+(\varepsilon-\mu)^2)$ in the integrand is large also increases, leading to an increase in $T_c$. The first two factors, which desire a decrease in $T_c$, dominate at small $\mu$, while the latter factor, which desires an increase in $T_c$, dominates at larger $\mu$. Together, these competing factors lead to the local minimum seen in Fig. \ref{fig:tcvsmu}.

We also see from Fig. \ref{fig:tcvsmu} that the value of $T_c$ at the minimum rapidly decreases as $f$ is increased. This is particularly prominent at larger values of $f$, as one can see from the inset of Fig. \ref{fig:tcvsmu}, where we take $f = 1.4$. Here, the minimum in $T_c$ is significantly more pronounced. As such, we find that our model effectively exhibits re-entrant superconductivity for larger values of $f$. That is, if one starts at large density and lowers the chemical potential, $T_c$ drops to zero, and then grows from zero to a constant as the chemical potential is further decreased and approaches zero. However, as is clear from the inset, the values of $T_c$ are so small that it is likely experimentally infeasible to observe this re-entrant superconductivity.

Finally, our results show that at $\mu =0$  there exists a critical $f^*$, above which superconductivity does not develop. We present our results for $T_c$ near $f^*$  in Fig. \ref{fig:scaling}.  We clearly see a power-law dependence of $T_c$ on $f^* -f$ which agrees with our analytical result,
$T_c \propto (f^*-f)^2$. We find $f^*\approx 4.9$ by performing a linear extrapolation of the calculated $(T_c/\omega_L)^{1/2}$ to 0. We emphasize that since $f^* >1$, there is a range of $f$ where superconductivity survives in the $\mu=0$ limit even when the pairing interaction is repulsive at all frequencies.

For larger values of $\bar{\mu}$ ($\bar{\mu}\gg 1$ or equivalently $\mu \gg \omega_L$), we return to the adiabatic regime, where only momenta near the Fermi level are relevant. In this case, we can track the behavior in $T_c$ by following the behavior in the coupling $\lambda(\bar{\mu}) = N(\bar{\mu}) V(\bar{\mu},\bar{\mu})$, which for our model is
\begin{equation}
    \lambda(\bar{\mu}) = \frac{1}{2\pi}\sqrt{\frac{\bar{\rho}}{\bar{\mu}}} \log(1+\pi \sqrt{\frac{\bar{\mu}}{\bar{\rho}}}).
\end{equation}

From this, we see that $\lambda(\bar{\mu})$ decays $\log(\bar{\mu})/\sqrt{\bar{\mu}}$ at $\bar{\mu} \gg \bar{\rho}$. Accordingly, for all $f>1$, we expect $T_c$ to go to zero as $\bar{\mu}$ is increased past some threshold $\bar{\mu}^*$, where the coupling $\lambda(\bar{\mu}^*)$ becomes too small to stabilize superconductivity. However, we expect this destruction of superconductivity to occur at extremely large $\bar{\mu}$ ($\bar{\mu} \gg \bar{\rho}$, where we have set $\bar{\rho}=\mathrm{Ry}/\omega_L = 136$) while the main focus of our work is on the low-density limit.

\subsubsection{Behavior of $\Delta_n(\varepsilon)$ with Matsubara Frequency}
\label{sec:gap_en_freq_dependence}
\begin{figure}
\centering
\includegraphics[width=\columnwidth]{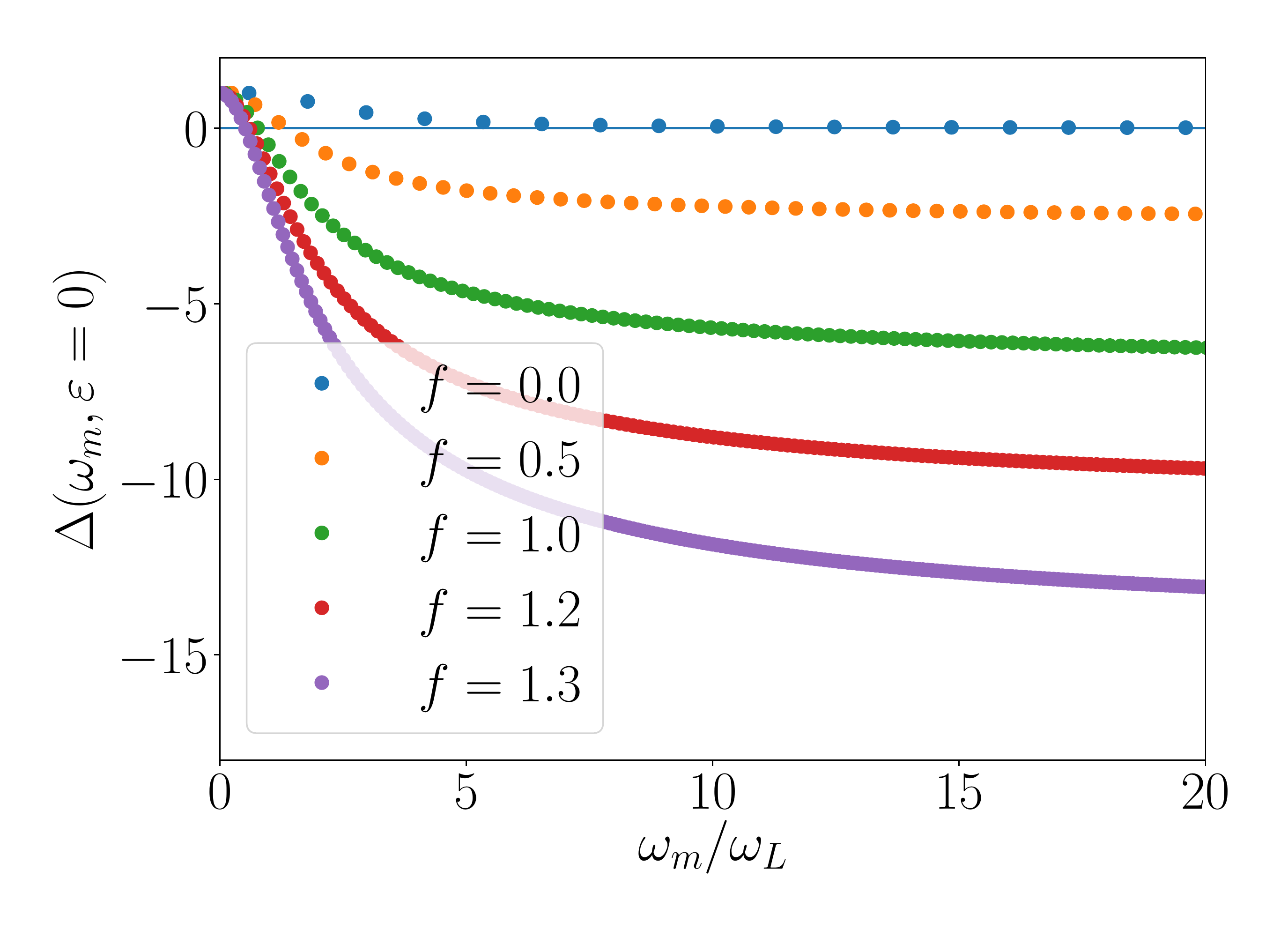}
\caption{\label{fig:gapvsfreq} $\Delta(\om,\varepsilon=0)$ for different values of $f$. We  have set $\mu = 10^{-5}\omega_L$.}
\end{figure}

We now turn to the behavior of the gap as a function of Matsubara frequency. The results are shown in Fig. \ref{fig:gapvsfreq}, where we have set $\mu = 10^{-5}\omega_L$. We find that for any $0<f< f^*$, $\Delta(\om)$ undergoes a sign change at some nonzero $\omega_0$.  The ratio $\Delta(\omega_m \gg \omega_L)/\Delta(\omega_m = 0)$ becomes more negative as $f$ increases. This fully agrees with the analytical result, presented in Fig. \ref{fig:gap_schematic}.

\begin{figure}
\includegraphics[width=\columnwidth]{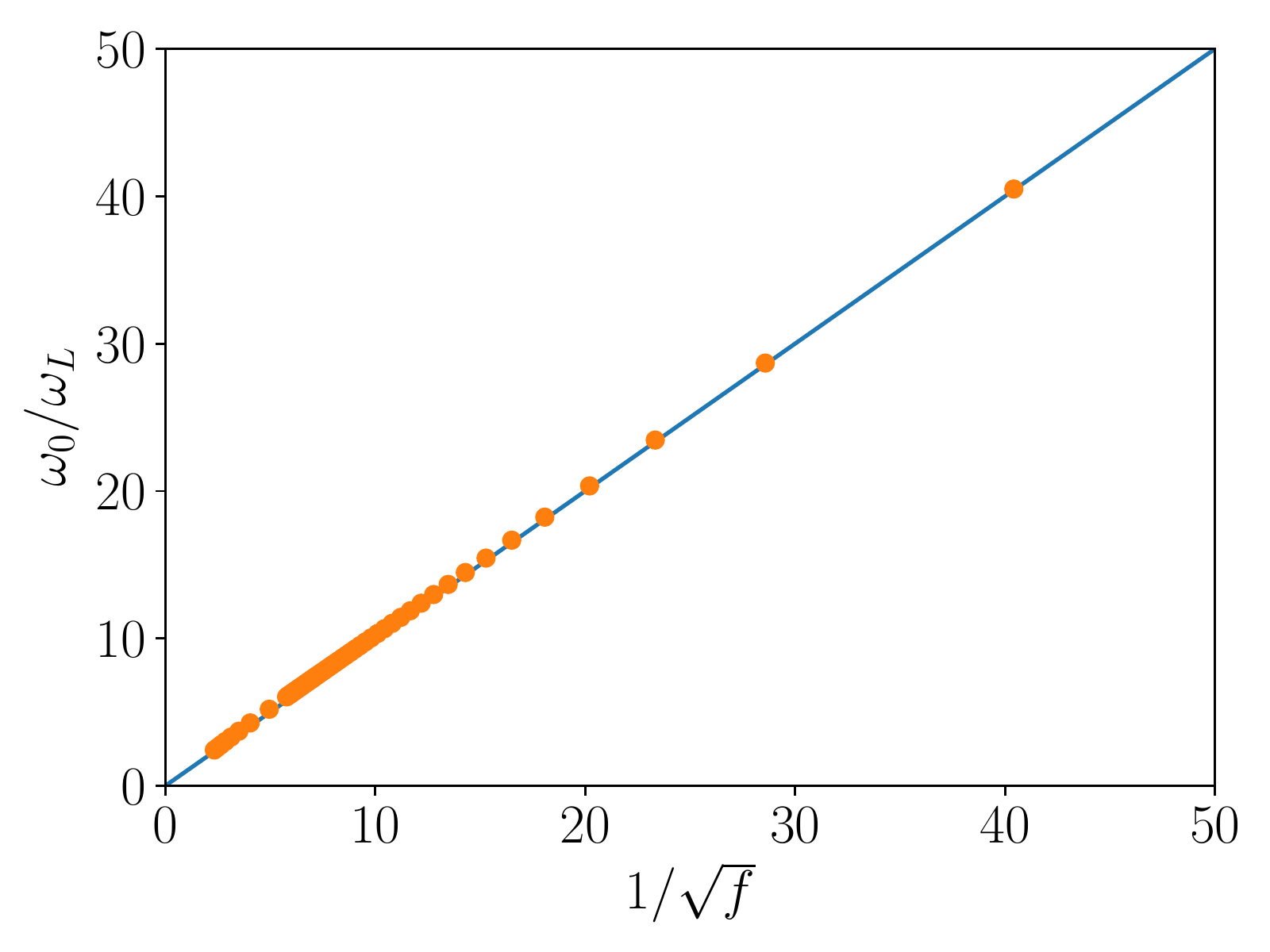}
\caption{\label{fig:gapzeros} The position of the gap node as a function of $1/\sqrt{f}$, for $\mu = 10^{-5}\omega_L$. }
\end{figure}

We now examine how $\omega_0$ behaves as a function of $f$. In Fig. \ref{fig:gapzeros} we show $\omega_0$ for small values of $f$. We find that the position of the node $\omega_0$ scales as $1/f^{1/2}$ at small $f$. This scaling is the same as at large density~\cite{Pimenov} and can be easily understood as the frequency dependence of the gap at large $\omega_m$ and small $f$ follows

\begin{equation}
    \Delta(\om) \propto \frac{\omega_L^2}{\om^2}
    -f.
\end{equation}

\begin{figure}
\includegraphics[width=\columnwidth]{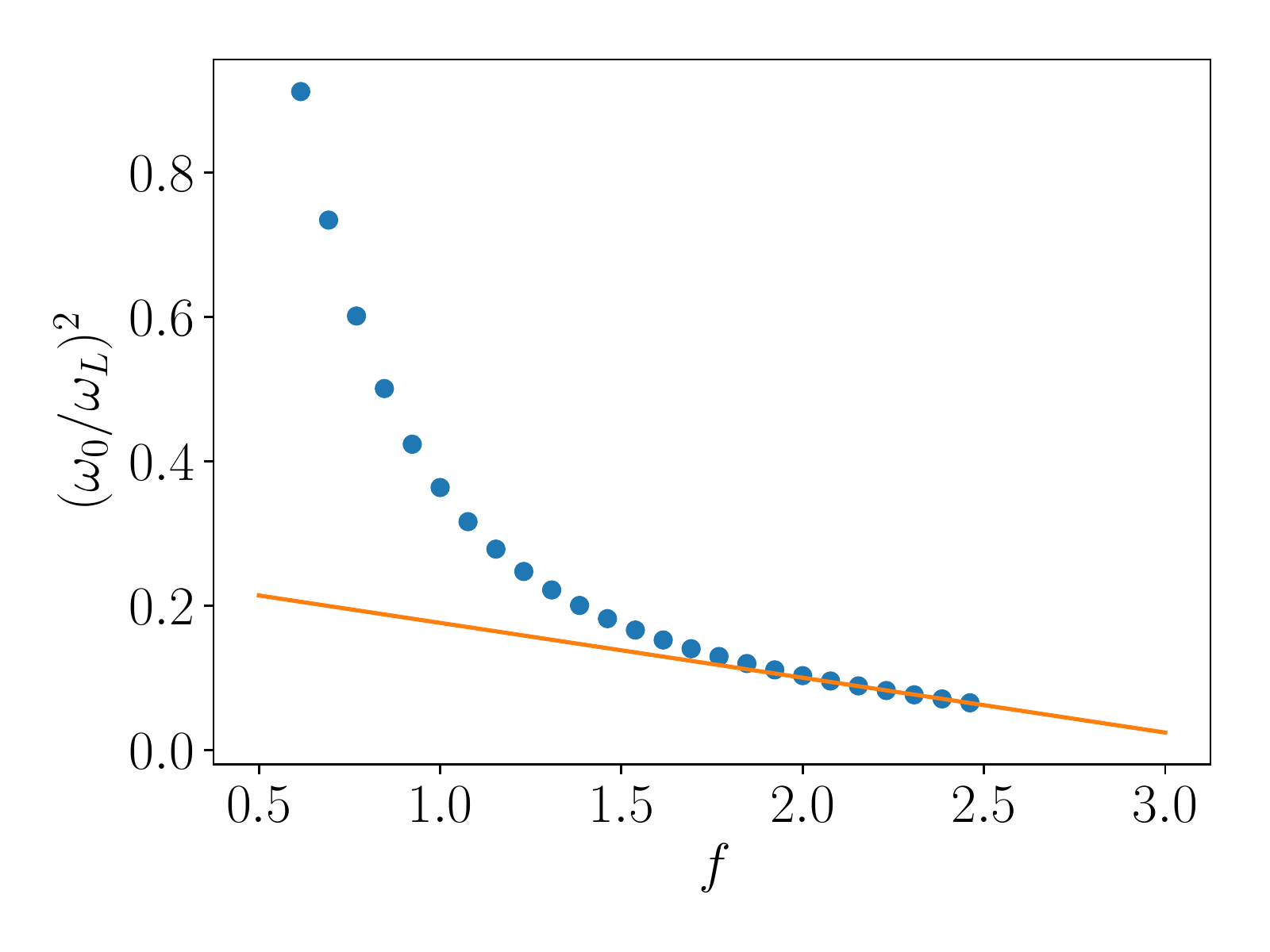}
\caption{\label{fig:gapf*} The $\mu = 0$ scaling of $\omega_0$ as $f$ approaches $f^*$. Overlaid is the line showing that $\omega_0 \propto (f^*-f)^{1/2}$ near $f=f^*$.}
\end{figure}

In Fig. \ref{fig:gapf*} we set $\mu=0$ and show how $\omega_0$ varies as $f$ approaches $f^*$. We find that $\omega_0$ decreases with $f$ nonlinearly, with the slope of $\omega_0^2 (f)$ decreasing with increasing $f$. The solid line in the plot is the fit to $\omega_0 \propto (f^*-f)^{1/2}$ that we obtained analytically in Eq. \ref{eq:gapeq_neqn_n7}. The fit is somewhat ambiguous as one needs more points closer to $f^*$. However, the agreement with our analytics is quite reasonable.

\subsubsection{Dependence of $\Delta_n(\varepsilon)$ on $\varepsilon$}

\begin{figure}
\includegraphics[width=\columnwidth]{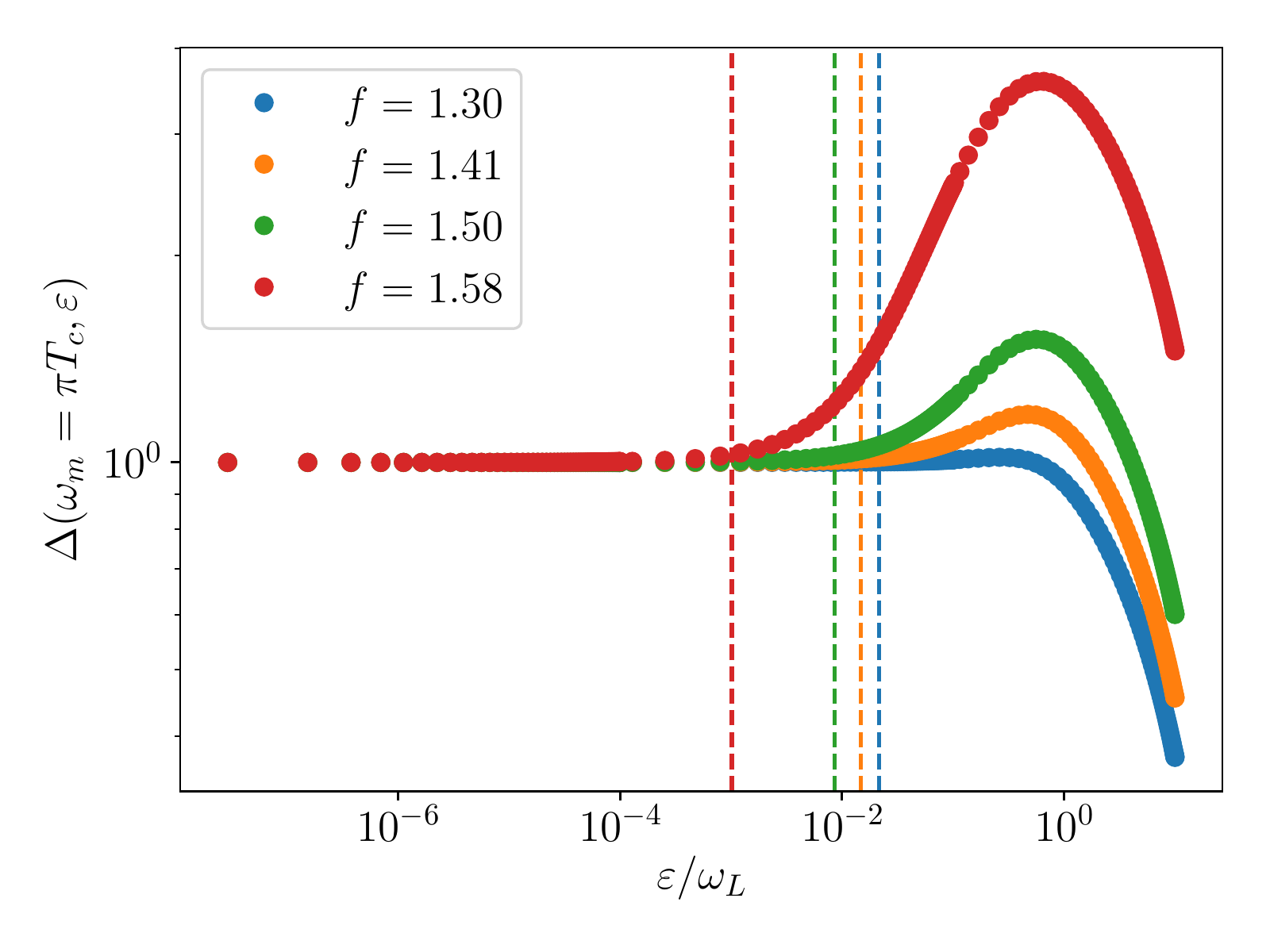}
\caption{\label{fig:gapvsen} The gap function $\Delta(\om = \pi T,\varepsilon)$ as a function of $\varepsilon$ for different values of $f$. The dashed lines show the values of $T_c/\omega_L$ for different $f$.  We have set $\mu = 10^{-5}\omega_L$.}
\end{figure}

We now turn to the behavior of the gap as a function of
$\varepsilon = k^2/2m$. We show the result in Fig. \ref{fig:gapvsen} for $\mu = 10^{-5}\omega_L$ for various values of $f$. Overlaid are dashed lines which delineate the values of $T_c/\omega_L$ for each value of $f$. From this plot, we see that there are essentially three regions of interest, (i) $\varepsilon \ll T_c$, (ii) $T_c \ll \varepsilon \ll \omega_L$, and (iii) $\varepsilon \gg \omega_L$. In region (i), the gap is essentially constant. In the intermediate region (ii) where $T_c \ll \varepsilon \ll \omega_L$, we find a smooth increase in the gap as a function of $\varepsilon$, which gets more pronounced with increasing $f$. Lastly, in region (iii) where $\varepsilon \gg \omega_L$, the gap decays as $B/\varepsilon$, where the constant $B$ grows with increasing $f$.
We note that the asymptotic behavior we see here agrees with our analytics, where we argued that the gap should be constant for small momenta and decay as $B/p^2$ (or equivalently $B/\varepsilon$) for large momenta.
Additionally, the behavior in the intermediate region also agrees with our analytical results, where we argued that there should be a peak in the momentum dependence of the gap, whose magnitude scales as $1/(f^*-f)$ at $\mu = 0$. However, we find numerically that the magnitude of this peak grows rather slowly as $f$ is increased. This may be due to a game of numbers.

We also note that $\Delta_n(\varepsilon)$ is smooth near the chemical potential.  We discuss this in more details in the next section.

\subsection{The Effect of the Momentum and Frequency Dependence of $\Pi(\Omega,q)$}
\label{sec:freq_dep}

\begin{figure}
\includegraphics[width=\columnwidth]{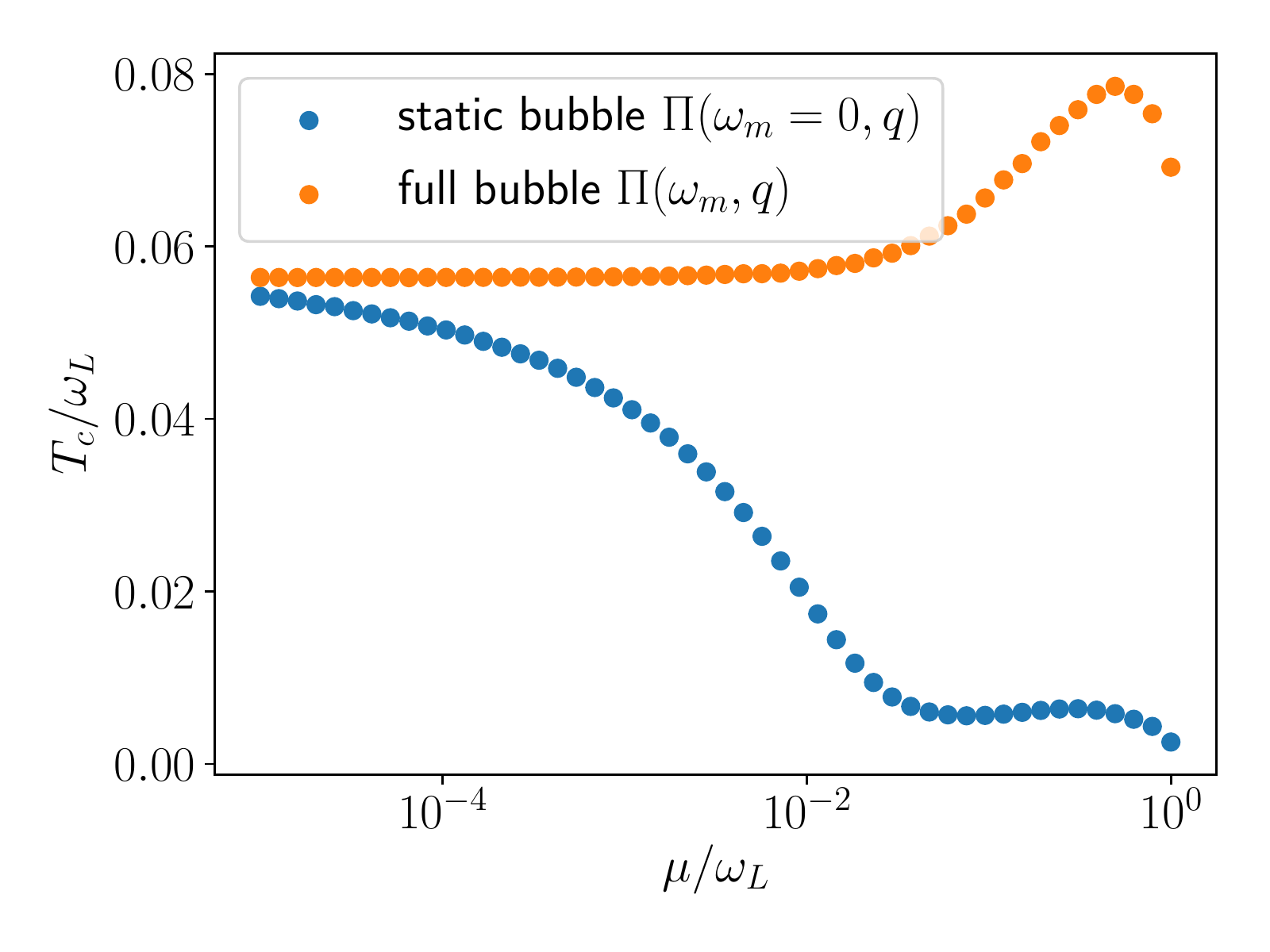}
\caption{\label{fig:bubble} $T_c$ vs $\mu$ for more general interactions. In the static bubble case, we include the momentum-dependence of the polarization bubble in the calculation, and set $f=1$. In the full bubble case, we use the interaction of Eq. \ref{eq:rpa}, setting $\omega_T = 0$ in the dielectric function; this is analogous to setting $f=1$. }
\end{figure}

To investigate the robustness of our results, we go beyond the extended Bardeen-Pines model and recalculate $T_c$ as a function of $\mu$, by (i) replacing $\kappa^2(\mu) = (-4\pi e^2\Pi(0,0))^{1/2}$ with $(-4\pi e^2\Pi(0,q))^{1/2}$ in Eq. \ref{eq:interaction}, and (ii) working with the full interaction Eq. \ref{eq:rpa}. In Case (i), we take $f=1$, while in Case (ii), we set  $\varepsilon_\infty = 1$ and $\omega_T = 0$ in the dielectric function $\varepsilon(\Omega)$; the latter is analogous to setting $f=1$. The results for these calculations are presented in Fig. \ref{fig:bubble}. In both cases, $T_c$ saturates to a nonzero value with decreasing $\mu$, in line with the results for the extended Bardeen-Pines model. This behavior is to be expected, since inclusion of the momentum and frequency dependence of $\Pi(\Omega,q)$ only weakens the screening.

\begin{figure}
\includegraphics[width=\columnwidth]{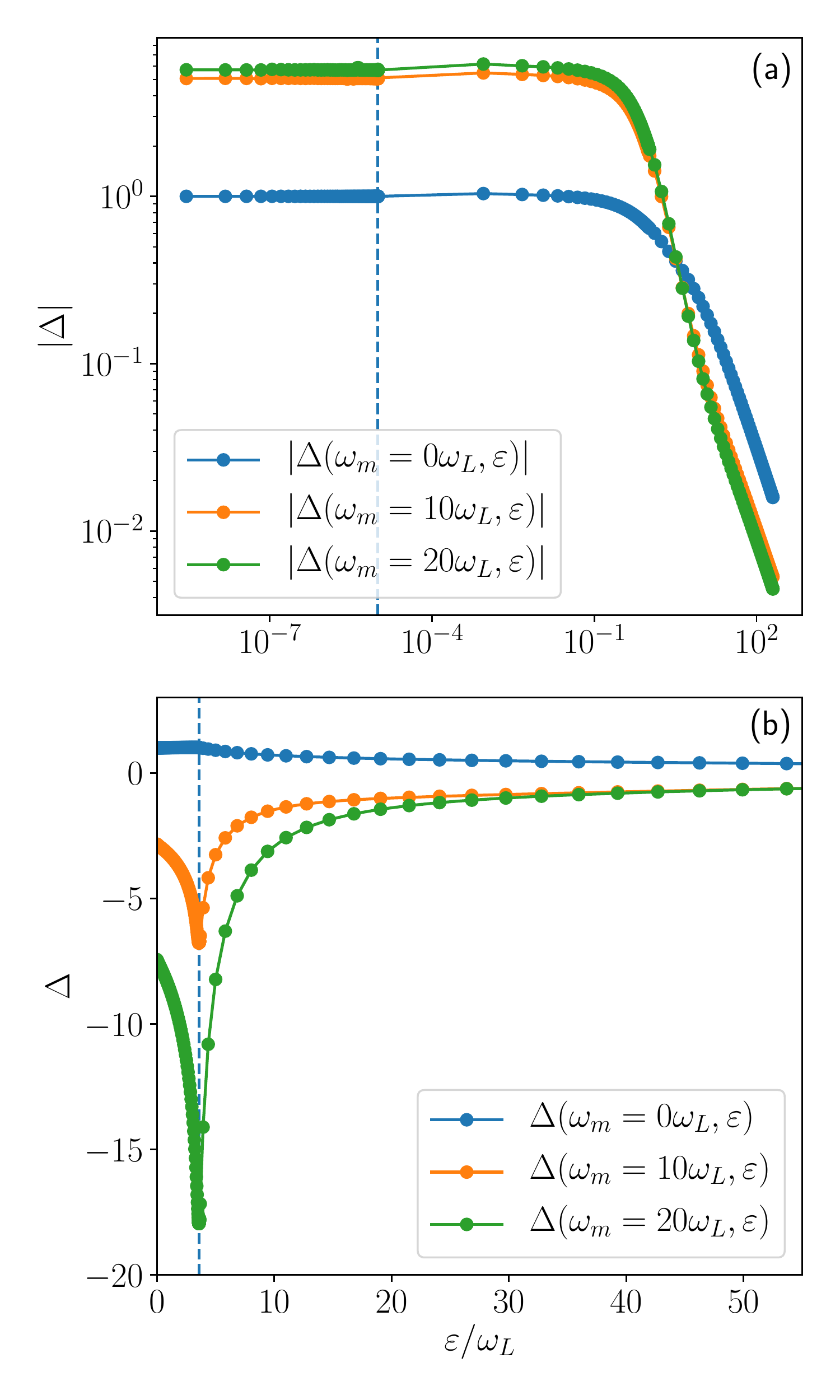}
\caption{The gap $\Delta_n(\varepsilon)$ as a function of energy $\varepsilon$ for various values of $\om$ in (a) the ultra-low density limit, and (b) the high-density limit, as obtained from the interaction with full dynamical screening. We have set $\omega_T = 0$ in these calculations, which is analogous to setting $f=1$. The chemical potentials ($\mu = 10^{-5}\omega_L$ in Panel (a),  $\mu = 3.59\omega_L$ in Panel (b)), are denoted with dotted lines.}
\label{fig:gap_kink}
\end{figure}

In Fig.\ref{fig:gap_kink}a we show the results for $\Delta_n (\varepsilon)$, obtained with the full $\Pi ( \Omega,q)$, as a function of energy $\varepsilon$ (not to be confused with the dielectric function $\varepsilon(\Omega)$) for various Matsubara frequencies, with $\mu = 10^{-5} \omega_L$.  We see that $\Delta_n (\varepsilon)$ is smooth at $\varepsilon \sim \mu$. This is consistent with the result that we obtained in the previous section for the extended Bardeen-Pines model.

In Fig. \ref{fig:gap_kink}b we present results of the same calculation, but at much larger density $\mu =3.59 \omega_L$.
 We see that at small Matsubara frequencies $\Delta_n (\varepsilon)$ is again smooth near the Fermi level, but at larger $\omega_n$ develops a strong dip at $\varepsilon = \mu$. Such a dip has been originally observed
by Takada \cite{Takada1992}. Richardson and Ashcroft \cite{Richardson1997} argued that it arises from the long-range nature of the Coulomb interaction and  holds when
$T_c/E_F \ll 1$ and one can linearize the dispersion around the Fermi level.

To understand this dip we set $Z_n(\varepsilon)=1$ and analyze the Eliashberg gap equation
\begin{equation}
\label{eq:sharp_dip_eliashberg}
\phi_n(\varepsilon) = - T_c \sum_m \int_0^\infty d\varepsilon' N(\varepsilon') V_{n-m}^\mathrm{s-wave}(\varepsilon,\varepsilon') \frac{\phi_m(\varepsilon')}{\om^2 + (\varepsilon'-\mu)^2}
\end{equation}
using
 \begin{equation}
\label{eq:full_V}
    V(\Omega,q) = \frac{4\pi e^2}{\varepsilon(\Omega)q^2-4\pi e^2\Pi(\Omega,q)}.
\end{equation}
We assume and then verify that in the limit of large Matsubara frequency, $\omega_n \to \infty$, relevant $\omega_m$ in the right-hand side of Eq. \ref{eq:sharp_dip_eliashberg} are finite.  The relevant bosonic $\Omega = \on-\om$ then approach $\infty$. Since the dynamical $\Pi (\Omega,q)$ vanishes for large $\Omega$, we have

\begin{equation}
   V_{n-m}^\mathrm{s-wave}(\varepsilon,\varepsilon') = V_{\infty}^\mathrm{s-wave}(\varepsilon_k,\varepsilon_q)
    = \frac{2\pi e^2}{ k q}\ln \frac{k+q}{|k-q|}.
\end{equation}
becomes purely static. Solving for the gap we then find that $|\phi_{\infty} (\varepsilon)|$  is logarithmically enhanced at $\varepsilon = \mu$:
\beq
\phi_{\infty} (\varepsilon = \mu) \sim \phi_{\infty} (\varepsilon \geq \mu) \ln^2\big(\frac{4\mu}{T_c}\big)
\eeq

The logarithmic singularity comes from the range near the Fermi surface, where $k \approx q$, which in turn arises from the long-range, unscreened behavior of the interaction at large $\on-\om$. We emphasize that this singularity in $\phi_\infty(\varepsilon)$ exists only in the high-density limit, and to obtain it one needs to include the full frequency dependence of the polarization bubble $\Pi(\Omega,q)$.

We also note that in this section we calculated $\Pi(\Omega,q)$ at $T = 0$. This is valid  when
$T_c \ll \mu(T_c)$ but is questionable as $\mu \to 0$. However, in light of the results presented here, we expect that including the temperature dependence of $\Pi(\Omega,q)$ should not qualitatively change our conclusions, since its inclusion would only serve to more quickly weaken the screening of the interaction.

\section{Conclusions}
\label{sec:conclusions}
In this work, we have studied the effect of a repulsive Coulomb interaction, on electron-phonon superconductivity in the low-density limit, the case of pairing interaction $V(\Omega,q) = 4\pi e^2/(q^2 + \kappa^2)\times(f- \omega^2_L/(\Omega^2 + \omega^2_L))$. Our results show that as for the $f=0$ case of pure electron-phonon attraction,  studied in Ref. \cite{Gastiasoro2019}, $T_c$ is enhanced as $\mu$ decreases, approaching a constant in the $\mu = 0$ limit. We find that the gap function changes sign at some Matsubara frequency $\omega_0$, reducing the effect of the repulsion and allowing $T_c$ to remain nonzero over some range of $1<f < f^*$, when the interaction is repulsive at all frequencies. As $f$ approaches $f^*$,  we find that both $T_c$ and $\omega_0$ approach zero as powers of $f^*-f$. This result, which we obtained both analytically and numerically, is in contrast to the behavior in the high-density limit, where $T_c$ vanishes exponentially in $f^*-f$.

Our results suggest that experimentally tuning the chemical potential should lead to substantial, observable variation in $\Delta(\omega_n)$, which can be observed in, e.g., ARPES experiments.

Lastly, we show that the behavior we find in $T_c$, namely that it stays nonzero when we take $\mu = 0$, continues to hold when we include dynamical screening of the interaction. Also, although in this work we focused on a 3D Galilean-invariant system, the behavior we find here should be relatively general and most likely continues to hold in two dimensions and for lattice systems.

In this work, we have not considered the possibility of other phases. Indeed, Wigner crystallization is also favored at low density. We leave study of the competition between superconductivity and other phases
to future work, noting only that a superconductor to Wigner-crystal phase transition has been previously proposed in the three-dimensional electron gas, where the electron-electron interaction is plasmon-mediated \cite{Takada1993}, and in twisted bilayer graphene \cite{Padhi2018}.

Another item for future study is the role of phase fluctuations. The transition temperature we calculate from the linearized Eliashberg equations is not the true superconducting transition temperature, but the onset temperature for pair formation. The superconducting transition temperature $T_c$ is defined as the onset of phase rigidity \cite{Pokrovsky1979} and in general should be smaller than the  onset temperature for the pairing. For obvious reasons we expect the effect of phase fluctuations to become progressively more relevant for quasi-2D systems.

\section*{Acknowledgements}
We thank M. Gastiasoro, R. Fernandes, D. Pimenov, J. Ruhman, Y. Wu and S. Zhang for useful discussions.
 The work was supported by the Office of Basic Energy Sciences US Department 791
of Energy under Award No. DE-SC0014402.  

\bibliography{main}

\end{document}